\documentclass[conference]{IEEEtran}
\usepackage{spconf}
\usepackage{tikz}
\usetikzlibrary{shapes.geometric, arrows, calc, positioning}

\usepackage{times}
\usepackage{helvet}
\usepackage{courier}
\usepackage{graphicx}
\usepackage{booktabs} 

\usepackage{url}
\usepackage[T1]{fontenc}    
\usepackage{booktabs}       
\usepackage{amsfonts}       
\usepackage{nicefrac}       
\usepackage{microtype}      
\usepackage[usenames,dvipsnames]{pstricks}
\usepackage{epsfig}
\usepackage{pst-grad} 
\usepackage{pst-plot} 
\usepackage{placeins} 
\usepackage{amsmath}
\usepackage{amssymb}
\usepackage{bbm}
\usepackage{theorem}
\usepackage{mathrsfs} 
\usepackage{euscript}
\usepackage{graphicx}

\usepackage[colorlinks=true,linkcolor=black,anchorcolor=black,citecolor=black,filecolor=black,menucolor=black,runcolor=black,urlcolor=blue]{hyperref}

\usepackage{float}
\usepackage{textcomp}
\usepackage{listings}
\usepackage{xcolor}
\usepackage{cases}
\usepackage{algorithm}
\usepackage{algorithmic}
\usepackage{multirow}
\usepackage[lofdepth,lotdepth]{subfig}
\graphicspath{{figures/}}
\renewcommand{\leq}{\ensuremath{\leqslant}}

\newcommand{\proj}{\mathrm{proj}}

\renewcommand{\epsilon}{\varepsilon}

\usepackage{blindtext} 

\title{Learning sparse auto-encoders for green AI image coding}
\name{%
Cyprien Gille $^{\ast}$, Frédéric Guyard $^{\dag}$,  Marc Antonini  $^{\ast}$, member IEEE, and Michel Barlaud $^{\ast}$, Fellow IEEE}
\address{$^{\ast}$ Université Côte d'Azur, I3S, CNRS, Sophia Antipolis, France\\
$^{\dag}$Orange Labs, Sophia Antipolis, France}

\begin{document}

\maketitle
\begin{abstract}
Recently, convolutional auto-encoders (CAE) were introduced for image coding. They achieved performance improvements over the state-of-the-art JPEG2000 method. However, these performances were obtained using massive CAEs featuring a large number of parameters and whose training required heavy computational power.\\
In this paper, we address the problem of lossy image compression using a CAE with a small memory footprint and low computational power usage. In order to overcome the computational cost issue, the majority of the literature uses Lagrangian proximal regularization methods, which are time consuming themselves.\\ 
In this work, we propose a constrained approach and a new structured sparse learning method. We design an algorithm and test it on three constraints: the classical $\ell_1$ constraint, the $\ell_{1,\infty}$ and the new $\ell_{1,1}$ constraint. Experimental results show that the $\ell_{1,1}$ constraint provides the best structured sparsity, resulting in a high reduction of memory and computational cost, with similar rate-distortion performance as with dense networks.
\end{abstract}


\section{Introduction}
Since Balle's \cite{Balle1} and Theis’ works \cite{Twitter} in 2017, most new lossy image coding methods use convolutional neural networks, such as convolutional autoencoders (CAE)
\cite{Balle2,Balle3,Mentzer,toderici}.  \\
CAEs are discriminating models that map feature points from a high dimensional space to points in a low dimensional latent space \cite{Hinton, Roweis}. They were introduced in the field of neural networks several years ago, their most efficient application at the time being dimensionality reduction and denoising \cite{stack-auto}. One of the main advantages of an autoencoder is the projection of the data in the low dimensional latent space : when a model properly learns to construct a latent space, it naturally identifies general, high-level relevant features. CAEs thus have the potential to address an increasing need for flexible lossy compression algorithms. In a lossy image coding scheme, the latent variable is losslessly compressed using entropy coding solutions, such as the well-known arithmetic coding algorithm. 

End-to-end training of a coding scheme reaches image coding performances competitive with JPEG 2000 (wavelet transform and bit plane coding) \cite{CAE-JPEG}. These are compelling results, as JPEG 2000 represents  the state-of-the-art for standardized image compression algorithms\footnote{{\color{blue} \url{https://jpeg.org/jpeg2000/index.html}}}.
Autoencoder-based methods specifically are becoming more and more effective : In a span of a few years, their performances have gone from JPEG to JPEG 2000. Considering the performances of these new CAEs for image coding, the JPEG standardization group has introduced the study of a new machine learning-based image coding standard, JPEG AI\footnote{{\color{blue}\url{https://jpeg.org/jpegai/index.html}}}.

Note that the performances of these CAEs are achieved at the cost of a high complexity and large memory usage.
In fact, energy consumption is the main bottleneck for running CAEs while respecting an energy footprint or carbon impact constraint \cite{AIgreen}, \cite{carbon}. Fortunately, it is known that CAEs are largely over-parameterized, and that in practice relatively few network weights are necessary to accurately learn image features.

Since 2016, numerous methods have been proposed in order to remove network weights ({\it weight sparsification}) during the training phase \cite{Tar2018}, \cite{Zho2016} \cite{Alv2016}.
These methods generally do produce sparse weight matrices, unfortunately with random sparse connectivity. To address this issue, many methods based on LASSO, group LASSO and exclusive LASSO were proposed \cite{Alv2016}, \cite{Hua2018}, \cite{Osw2016} in order to simultaneously sparsify neurons and enforce parameter sharing. 
However, all proximal regularization methods quoted above require the computation of the Lasso path, which is time consuming \cite{hrtzER}.

In order to deal with this issue, we proposed instead a constrained approach in \cite{BBCF}, where the constraint is directly related to the number of zero-weights.
In \cite{BG20}, we designed an algorithm for the sparsification of linear fully connected layers in neural networks with three constraints in mind: the classical $\ell_1$ constraint, the $\ell_{1,\infty}$ constraint and the structured \cite{ICASSP} $\ell_{1,1}$ constraint.

In this work, we extend the aforementioned approach to a CAE in the context of image coding in order to reduce its computational footprint while keeping the best rate-distortion trade-off as possible. We detail in section \ref{Sparse} the constraint approach we developed to sparsify the CAE network. In section \ref{exp}, we present the first experimental results. Finally, section \ref{Conclusion} concludes the paper and provides some perspectives.

\section{Learning a sparse autoencoder using a structured constraint}
\label{Sparse}
\begin{figure}
    \centering
    \includegraphics[width=0.49\textwidth,height=5cm]{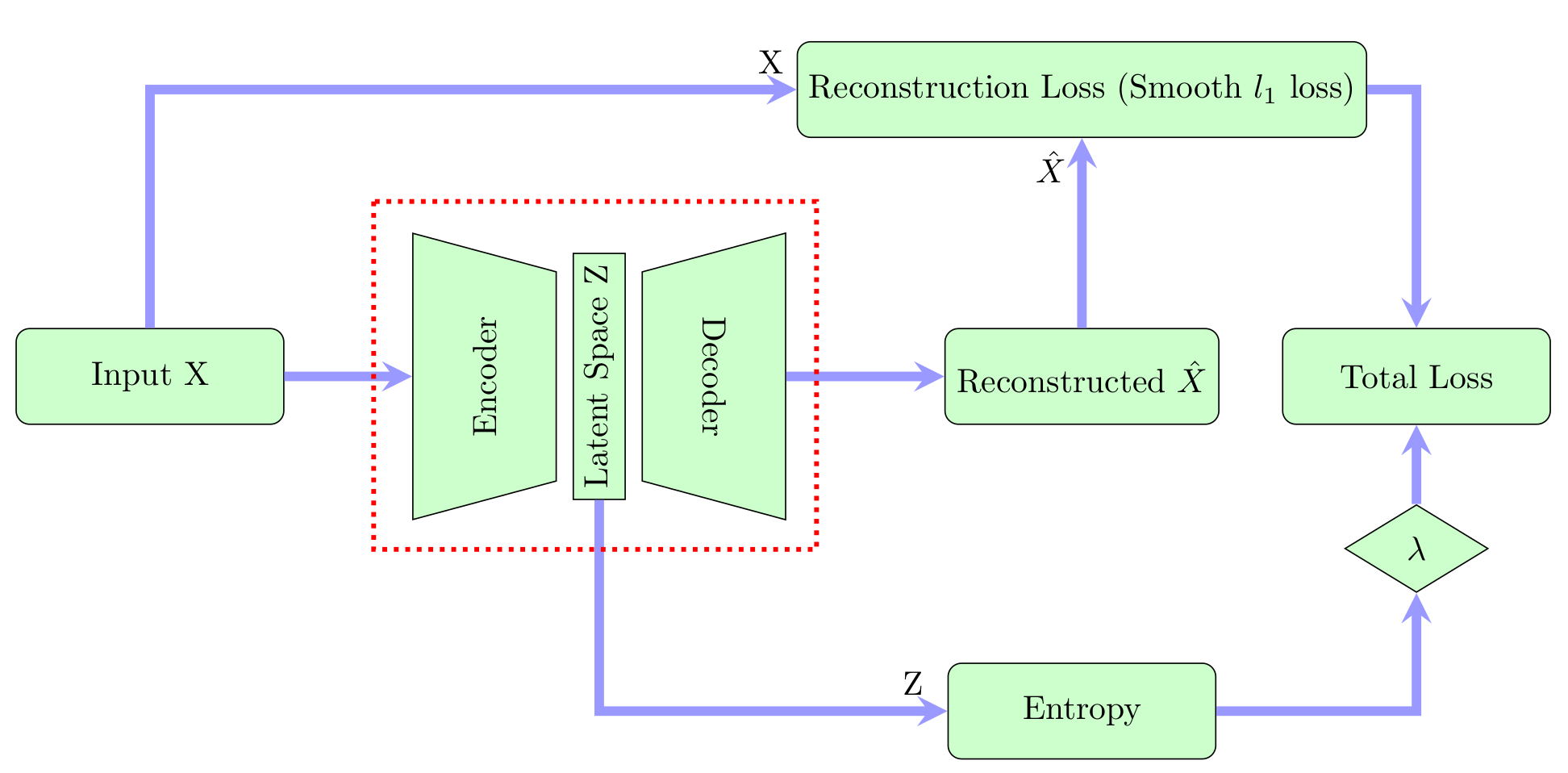}
    \caption{Entropy-Distorsion scheme with a CAE}
    \label{Auto}
\end{figure}

Figure \ref{Auto} shows the architecture of a CAE network where $X$ is the input data, $Z$ the latent variable and $\widehat{X}$ the reconstructed data. In the following, let us call $W$ the weights of the CAE. 

The goal is to compute the set of weights of the CAE $W$ minimizing the total loss for a given training set. The total loss corresponds to the classical trade-off between entropy and distortion, and is thus a function of the entropy of the latent space $Z$ and of the reconstruction error between $X$ and $\hat{X}$. It can then be modified to achieve weight sparsification with a regularization term.

Let us recall the classical Lagrangian regularization approach as the following:  
\begin{equation}
Loss(W) = \lambda \cdot \mathcal{H} ( Z)+ \psi (\widehat{X} -X) + \alpha  \|W\|_1^1,
\label{alpha-loss}
\end{equation}
where $\mathcal{H}(Z)$ is the entropy of the latent variable distribution and $\psi$ is the reconstruction loss, for which we use the robust Smooth $\ell_1$ (Huber) loss. 
However, the main issue of (\ref{alpha-loss}) is that the computation of parameter $\alpha$ using the Lasso path is computationally expensive \cite{hrtzER}. In order to deal with this issue, we propose to minimize the following constrained approach instead:
\begin{equation}
Loss(W) = \lambda \cdot \mathcal{H}(Z) + \psi (\widehat{X} -X) \text{ s.t. } \|W\|_1^1 \leq  \eta.
\label{constraint}
\end{equation}
with $\eta$ being the projection radius.

The main difference with the criterion proposed in \cite{Twitter} is the introduction of the constraint on the weights $W$ to sparsify the neural network. Low values of $\eta$ imply high sparsity of the network.

The classical Group LASSO consists of using the $\ell_{2,1}$ norm for the constraint on $W$ (instead of the $\ell_{1,1}$ norm as proposed in \cite{ICASSP}). However, the $\ell_{2,1}$ norm does not induce a structured sparsity of the network \cite{barlaud2019}, which leads to negative effects on performance when attempting to reduce the computational cost.

The projection using the $\ell_{1,1}$ norm is computed by Algorithm  \ref{algo1-proj_L21}.
We first compute the radius $t_i$, and then project the rows using $t_i$ as the adaptive constraint for $\ell_1$. Note that in the case of a CAE, contrarily to fully connected networks, $W$ originally corresponds to a tensor instead of a matrix. We thus need to flatten the inside dimensions of the weight tensors to turn them into two-dimensional arrays.
Algorithm \ref{algo1-proj_L21} requires the projection of the vector $(\|v_i\|_1)_{i=1}^d$ on the $\ell_1$ ball  of radius $\eta$ whose complexity is only $O (d)$ \cite{condat,perez19}.
We then run the double descent Algorithm \ref{algoglobal} \cite{double,Lottery} where instead of the weight thresholding done by state-of-the-art algorithms, we use our $\ell_{1,1} $ projection from Algorithm \ref{algo1-proj_L21}.

\begin{algorithm}
\begin{algorithmic}
\STATE \textbf{Input:} $V,\eta$
  \FOR{$i = 1,\dots,d$}
  \STATE{$t_i :=\proj_{\ell_1}((\|v_j\|_1)_{j=1}^l,\eta)_i$}
  \STATE{$w_i := \proj_{\ell_1}(v_i,t_i)$} 
  \ENDFOR
\STATE \textbf{Output:} $W$
\end{algorithmic}
\caption{Projection of the $l \times d$ matrix $V$ onto the $\ell_{1,1}$-ball of radius $\eta$. $\proj_{\ell_1}(v,\eta)$ is the projection of $v$ on the $\ell_1$-ball of radius $\eta$ }
\label{algo1-proj_L21}
\end{algorithm}

\begin{algorithm}
\begin{algorithmic}
\STATE \textit{\# First descent}
\STATE \textbf{Input:} $W_{init},\gamma,\eta$ 
\FOR{$n = 1,\dots,N$}
  \STATE $W \leftarrow A(W,\gamma, \nabla \phi (W) )$
\ENDFOR
\STATE \textit{\# Projection}
  \FOR{$i = 1,\dots,d$} 
  \STATE{$t_i := proj_{\ell_1}((\|v_j\|_1)_{j=1}^l,\eta)_i$}
  \STATE{$w_i := proj_{\ell_1}(v_i,t_i)$} 
  \ENDFOR
\STATE $(M_0)_{ij} := \mathbbm{1}_{x \ne 0} (w_{ij})$
\STATE \textbf{Output:} $M_0$
\STATE \textit{\# Second descent}
\STATE \textbf{Input:} $W_{init}, M_0, \gamma$
\FOR{$n = 1,\dots,N$}
  \STATE $W \leftarrow A(W,\gamma, \nabla \phi (W,M_0) )$
\ENDFOR
\STATE \textbf{Output:} $W$
\end{algorithmic}
\caption{Double descent algorithm. $\phi$ is the total loss as defined in (\ref{constraint}), $\nabla \phi (W,M_0)$ is the gradient masked by the binary mask $M_0$, $A$ is the Adam optimizer, $N$ is the total number of epochs and $\gamma$ is the learning rate.}
\label{algoglobal}
\end{algorithm}

\section{Experimental results}
\label{exp}
\subsection{Settings}
The proposed method was implemented in PyTorch using the python code implementation of a convolutionnal auto-encoder proposed in \cite{Twitter}\footnote{{\color{blue}\url{https://github.com/alexandru-dinu/cae}}}.
Note that the classical computational cost measure evaluates FLOPS (floating point operations per second), in which additions and multiplications are counted separately. However, a lot of modern hardware can compute the multiply-add operation in a single instruction. Therefore, we instead use {\bf MACCs} (multiply-accumulate operations) as our computational cost measure (multiplication and addition are counted as a single instruction\footnote{{\color{blue}\url{https://machinethink.net/blog/how-fast-is-my-model/}}}).

We trained the compressive autoencoder on 473 $2048 \times 2048$ images obtained from Flickr\footnote{{\color{blue}\url{https://github.com/CyprienGille/flickr-compression-dataset}}}, divided into $128 \times 128$ patches. We use the 24-image Kodak PhotoCD dataset for testing \footnote{{\color{blue}\url{http://www.r0k.us/graphics/kodak/}}}. All models were trained using 8 cores of an AMD EPYC 7313 CPU, 128GB of RAM and an NVIDIA A100 GPU (40GiB of HMB2e memory, 1.5TB/s of bandwidth, 432 Tensor Cores). Performing 100 Epochs of training takes about 5 hours.

We choose as our baseline a CAE network trained using the classical Adam optimizer in PyTorch, and compared its performance (relative MACCs and loss as a function of sparsity, PSNR and Mean SSIM as a function of the bitrate) to our masked gradient optimizer with $\ell_1$, $\ell_{1,1}$ and $\ell_{1,\infty}$ constraints. For the $\ell_{1, \infty}$ projection, we implemented the "Active Set" method from \cite{Bejar}. For the PSNR function, we used its implementation in CompressAI \cite{begaint2020compressai}\footnote{\url{https://github.com/InterDigitalInc/CompressAI}}. Note that \textit{MSSIM} refers to the Mean Structural SIMilarity, as introduced in \cite{MSSIM}. Considering that the high image quality and low distortions are difficult to assess within an article, we provide here a link to our website (\url{https://www.i3s.unice.fr/~barlaud/Demo-CAE.html}) so that the reader can download the images and evaluate their quality on a high definition screen.



From now on, we use $S$ to denote the encoder sparsity proportion, i.e. the ratio of zero-weights over the total number of weights in the encoder.

\subsection{Sparsification of the encoder}
In this section, we only sparsify the encoder layers of the CAE. This is a practical case, where the power and memory limitations apply mostly to the sender as is the case for satellites \cite{rennes,satellite-denoising-ieee,johnston-efficient}, drones, or cameras used to report from an isolated country.

\begin{figure}[!h]
    \centering
    \includegraphics[width=0.49\textwidth,height=5.cm]{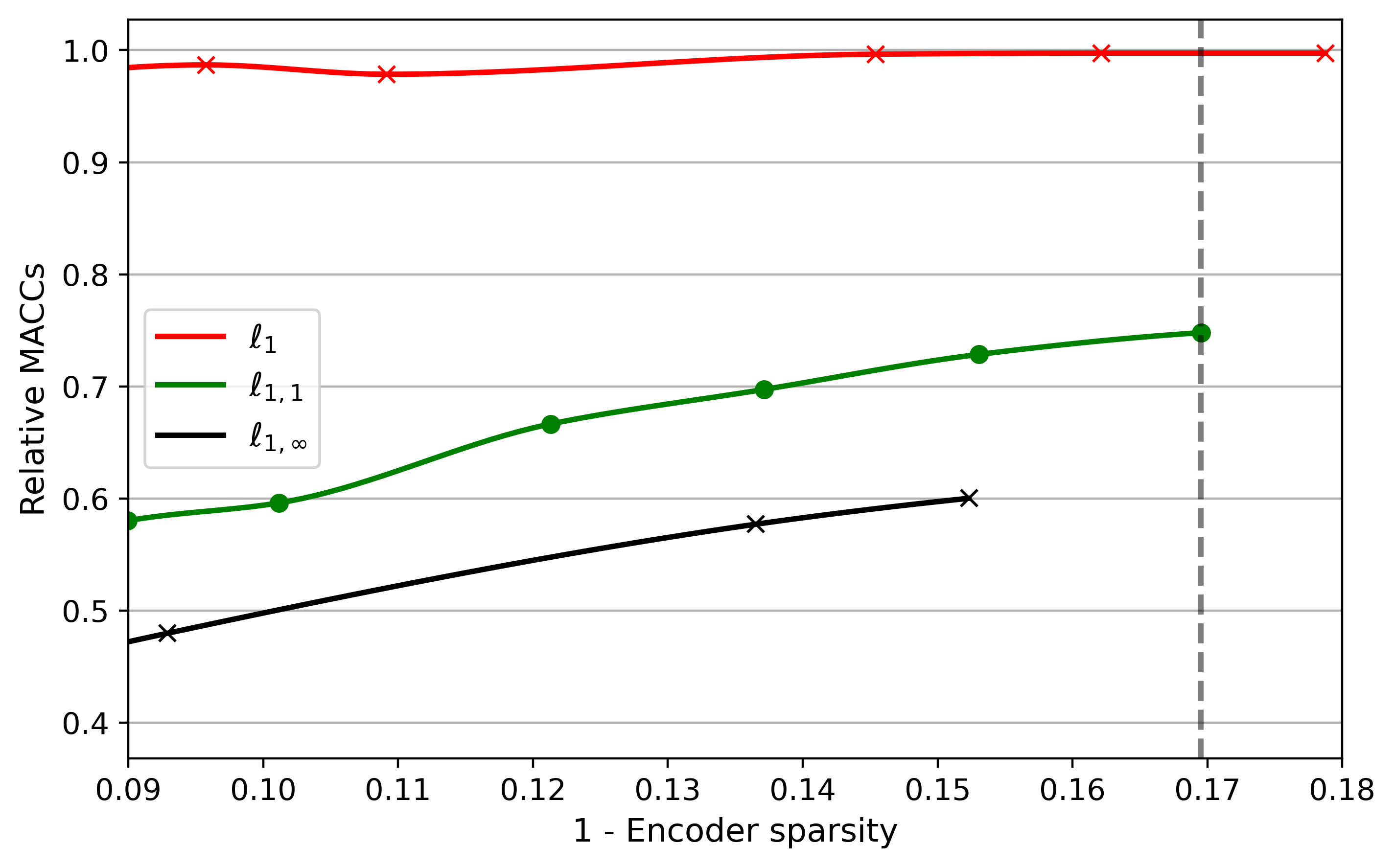}
    \caption{Projection on encoder layers : Relative MACCs computational cost as a function of the density.}
    \label{rmaccs-encodproj}
\end{figure}

Figure \ref{rmaccs-encodproj} displays the relative number of MACCs with respect to the aforementioned baseline of a non-sparsified network, as a function of the density $1-S$. The $\ell_{1}$ constraint does not provide any computational cost improvement while the $\ell_{1,1}$ constraint significantly reduces MACCs. This is due to the fact that the $\ell_{1,1}$ constraint (contrarily to $\ell_{1}$) creates a structured sparsity \cite{barlaud2019}, setting to zero groups of neighbouring weights, inhibiting filters and thus pruning operations off the CAE. The $\ell_{1, \infty}$ constraint reduces MACCs even more, but comes with a reduction of the performance of the network.

Let us now define the relative PSNR loss with respect to the reference without projection as:
\begin{center}
   $ 10\times(log_{10}(MSE_{ref}) -log_{10}(MSE_{L})) \ \text{in dB}$   
\end{center}
for a constraint $L=\ell_1$, $L=\ell_{1,1}$, or $L=\ell_{1,\infty}$.\\
Table \ref{loss-encodproj} shows the relative loss of the different models for a value of the sparsity $S$ around $83\%$. The table shows that the $\ell_{1,1}$ constraint leads to a slightly higher loss than $\ell_1$. However, this slight decrease in performance comes with a significant decrease in computational cost ($30\%$ less MACCs for a sparsity of $86\%$), as shown in Figure \ref{rmaccs-encodproj}.\\

\begin{figure}[!h]
    \centering
    \includegraphics[width=0.49\textwidth,height=5.cm]{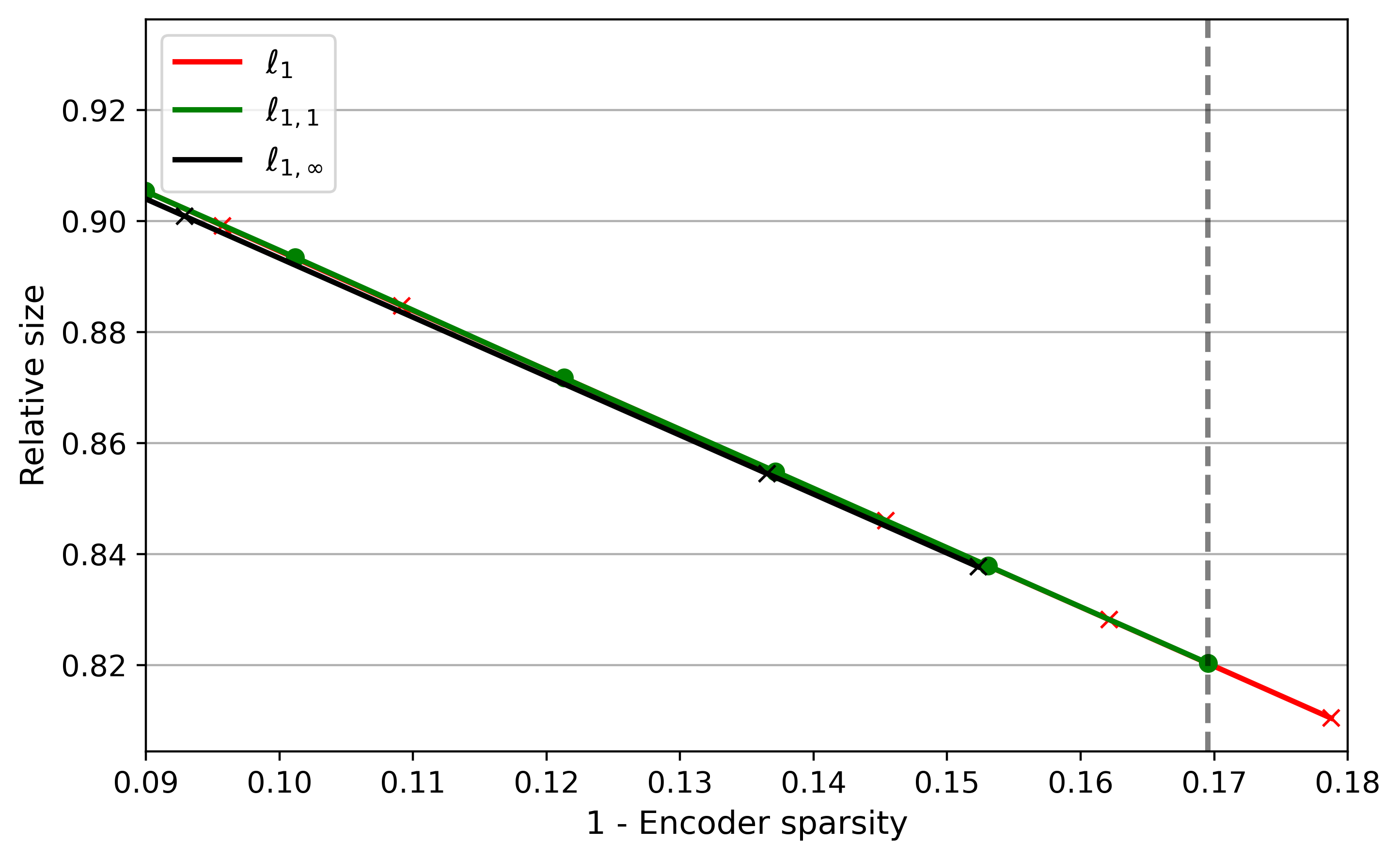}
    \caption{Projection on encoder layers : Relative memory reduction as a function of the density.}
    \label{size-encodproj}
\end{figure}
    
\begin{table}[htp!]
    \centering
    \caption{Projection on encoder layers : Sparsity, MACCs, Memory reduction and relative loss}
    \begin{tabular}{|c|c|c|c|}
        \hline
        Constraint  & $\ell_{1} $  & $\ell_{1,1} $  & $\ell_{1,\infty}$  \\
        \hline
        $S$ (\%) & 82.12 & 83.05  & 84.77\\
        \hline
        MACCs reduction $ \% $  & 0 & 27 & 40\\
        \hline
        Memory reduction $ \% $  & 81 & 82 & 84\\
        \hline
        Relative Loss (dB)  & -1.15 & -1.2 & -1.7\\
        \hline
    \end{tabular}
    \label{loss-encodproj}
\end{table}
    
We then display the bitrate-distortion curves for the CAEs from Table \ref{loss-encodproj}.

\begin{figure}[!ht]
    \centering
    \includegraphics[width=0.49\textwidth,height=5.cm]{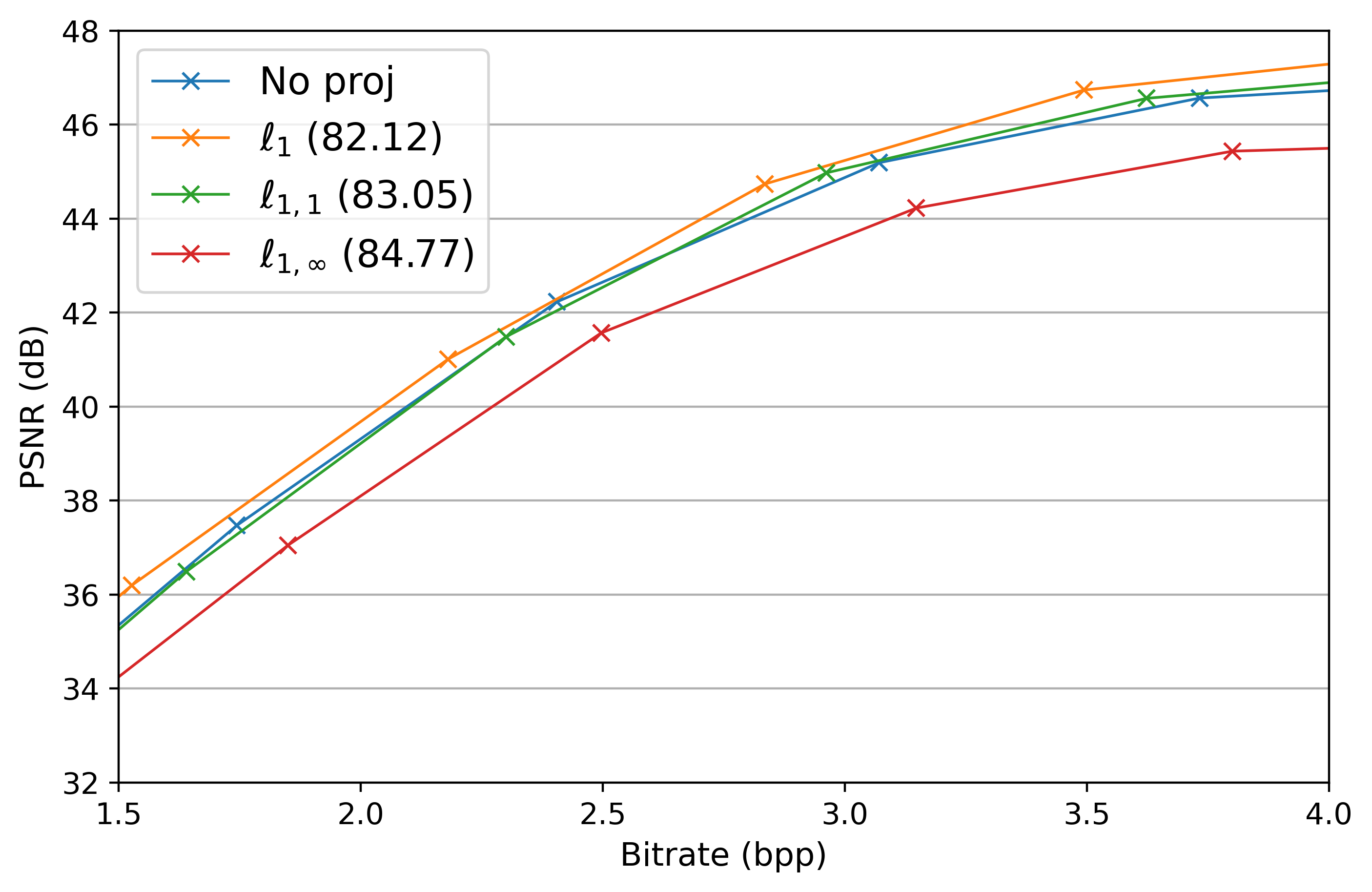}
    \caption{Projection on encoder layers : PSNR as a function of the bitrate for decoded Kodak test images}
    \label{PSNR-encodproj}
\end{figure}

\begin{figure}[!ht]
    \centering
    \includegraphics[width=0.49\textwidth,height=5.cm]{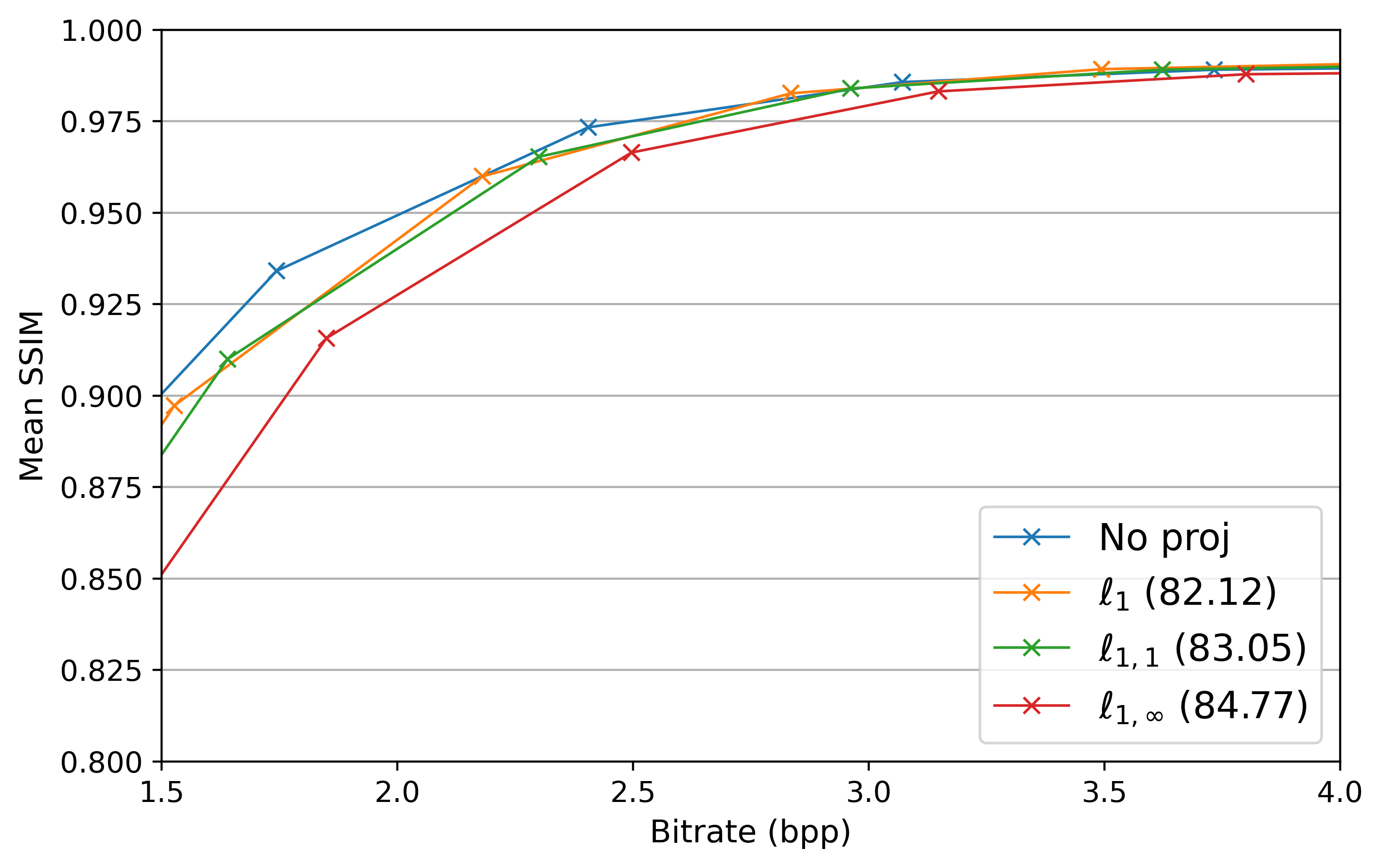}
    \caption{Projection on encoder layers : Mean SSIM as a function of the bitrate for decoded Kodak test images}
    \label{SSIM-encodproj}
\end{figure}

The figures \ref{PSNR-encodproj} and \ref{SSIM-encodproj} show a slight PSNR loss of less than $1 dB$, and similarly close MSSIM scores.
This loss translates perceptually into a slight reinforcement of the image grain, which is more noticeable for projection $\ell_{1,\infty}$.

\begin{figure}[!ht]
    \centering
    \subfloat[\centering Original]{%
    \label{img-orig-encproj}%
    \includegraphics[width=0.24\textwidth,height=4.cm]{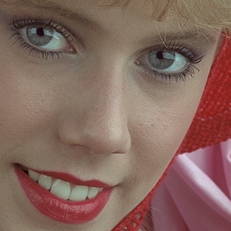}}%
    \enspace
    \subfloat[\centering No Projection]{%
    \label{img-initial-encproj}%
    \includegraphics[width=0.24\textwidth,height=4.cm]{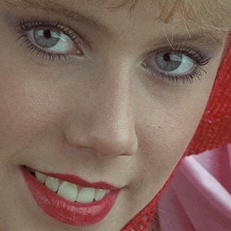}}%
    \\
    \subfloat[\centering $\ell_{1,1}$, $S=87.87$, $RM=34$]{%
    \label{img-l11-encproj}%
    \includegraphics[width=0.24\textwidth,height=4.cm]{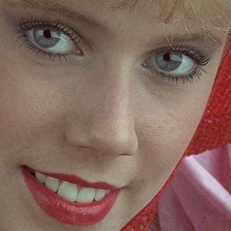}}%
    \enspace
    \subfloat[\centering $\ell_{1,\infty}$, $S=86.35$, $RM=42$]{%
    \label{img-l1inf-encproj}%
    \includegraphics[width=0.24\textwidth,height=4.cm]{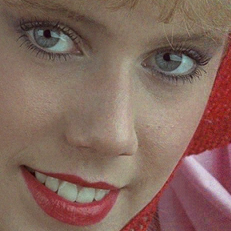}}%
    \caption{Projection of the encoder layers : Comparison of reconstructed test images for different models. $S$ is the sparsity, $RM$ refers to the percentage of MACCS reduced by the projections. Bitrates around $2.25bpp$.}
\end{figure}


\subsection{Sparsification of all layers}

\begin{table}[ht!]
    \centering
    \caption{Projection on all layers: Sparsity, MACCs and relative loss}
    \begin{tabular}{|c|c|c|c|}
        \hline
        Projection   & $\ell_{1} $  & $\ell_{1,1} $  & $\ell_{1,\infty}$ \\
        \hline
        Encoder $S$ (\%) & 83.28 & 84.21  & 85.43\\
        \hline
        MACCs Reduction $ \% $  & 0 & 30 & 47\\
        \hline
        Memory Reduction $ \% $  & 79 & 80 & 83\\
        \hline
        Relative Loss (dB)  & -4.40 & -4.48 & -11.88\\
        \hline
    \end{tabular}
    \label{loss-fullproj}
\end{table}

Table \ref{loss-fullproj} shows the relative loss of the different models for a given value of $S$ around $84\%$. The table shows that the $\ell_{1,1}$ and $\ell_1$ constraints perform very similarly, with $\ell_{1,1}$ still being the only one of those two constraints to offer MACCs reduction. The 30\% difference of computational cost reduction between $\ell_{1, \infty}$ and $\ell_{1,1}$ has been reduced to only $17\%$, with the noticeable loss in performance for $\ell_{1, \infty}$ remaining.

\begin{figure}[!ht]
    \centering
    \includegraphics[width=0.49\textwidth,height=5.cm]{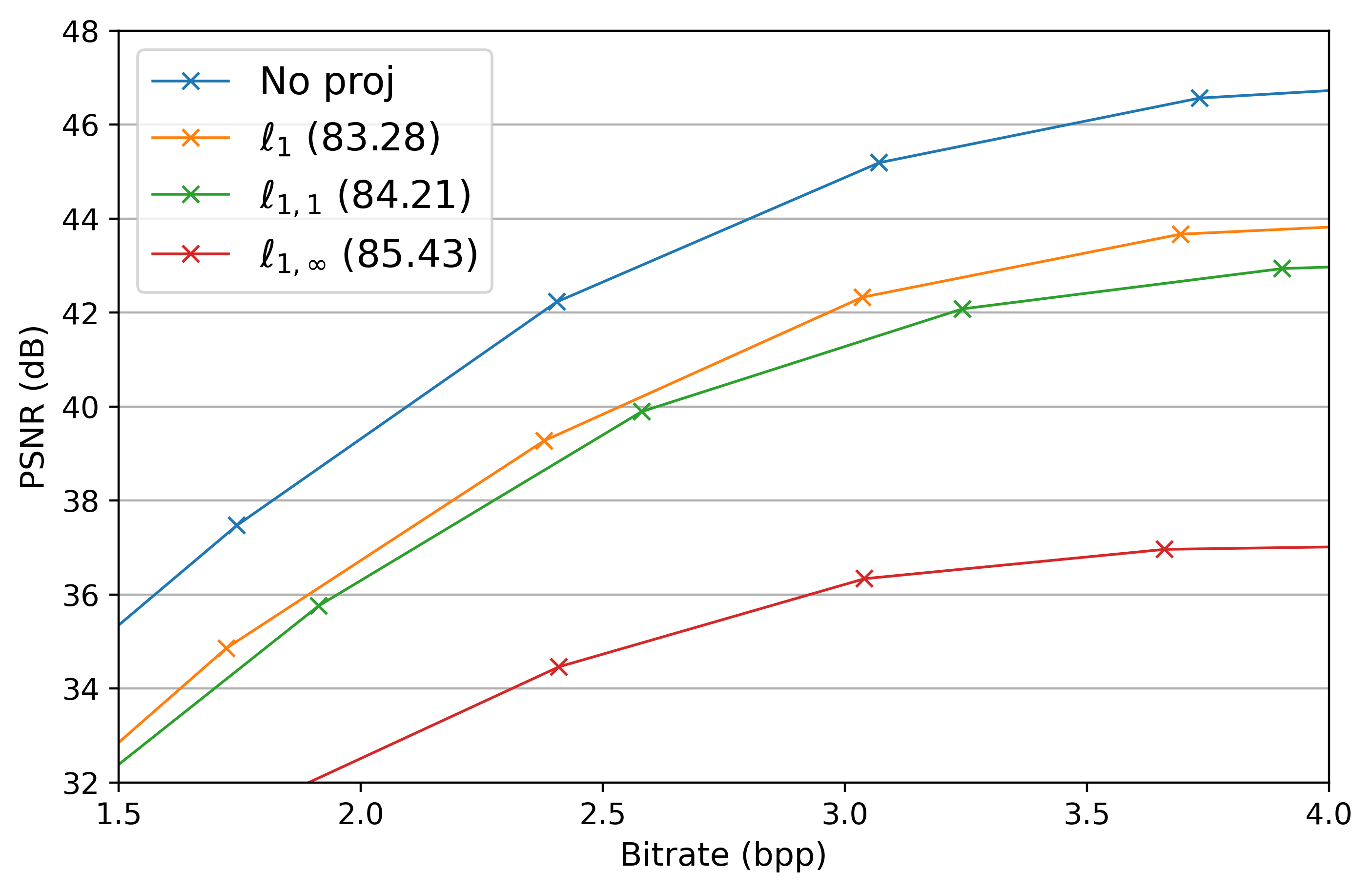}
    \caption{Projection on all layers : PSNR as a function of the bitrate for decoded Kodak test images}
    \label{PSNR-fullproj}
\end{figure}

\begin{figure}[!ht]
    \centering
    \includegraphics[width=0.49\textwidth,height=5.cm]{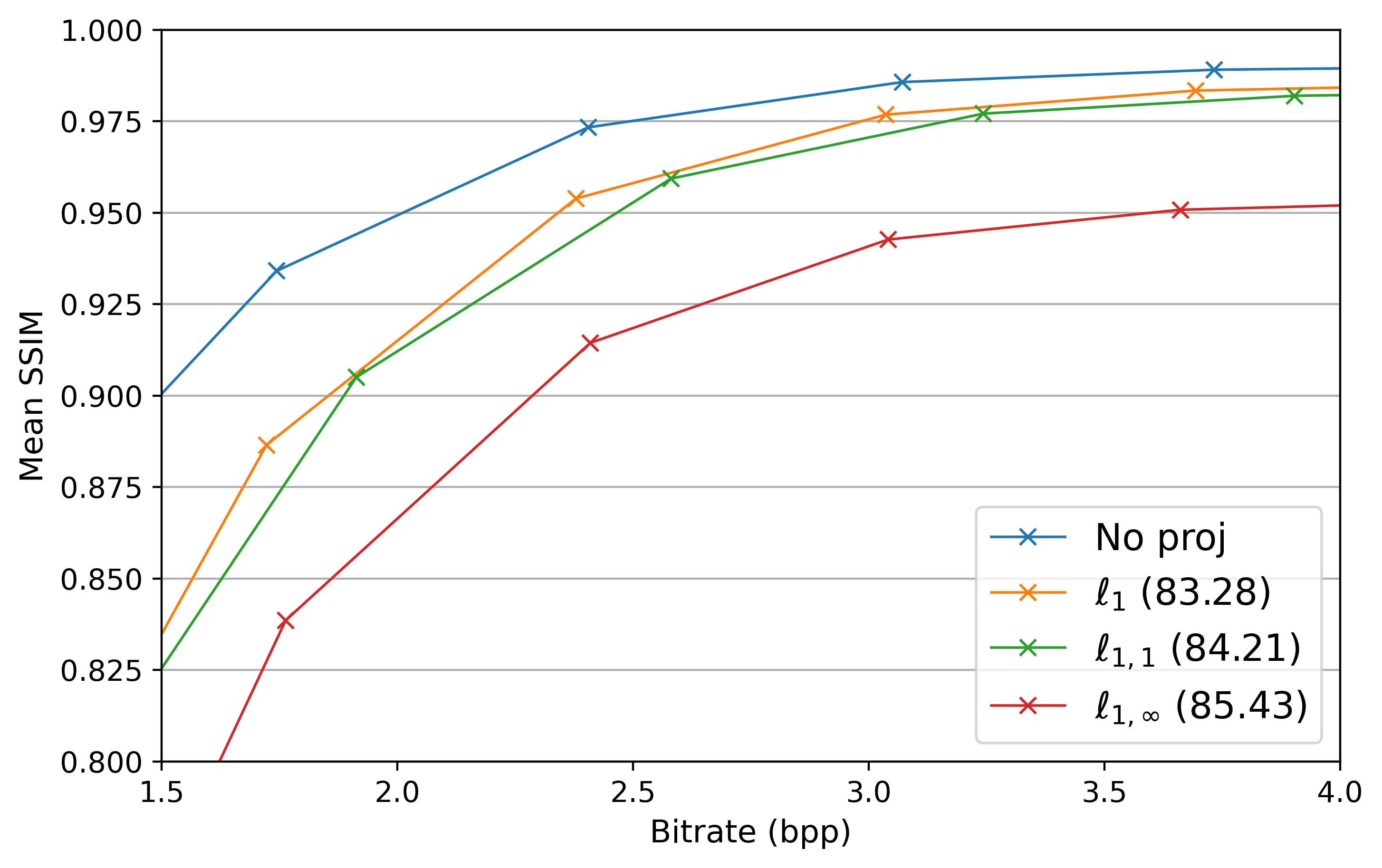}
    \caption{Projection on all layers : Mean SSIM as a function of the bitrate for decoded Kodak test images}
    \label{SSIM-fullproj}
\end{figure}

Figures \ref{PSNR-fullproj} and \ref{SSIM-fullproj} show that despite the sizeable reduction in computational cost provided by the $\ell_{1,1}$ constraint, the reconstructed images are only around $3.5$dB of PSNR worse than the ones obtained without projection. The $\ell_{1, \infty}$ constraint still performs markedly worse than the other projection.

\begin{figure}[!h]
    \centering
    \subfloat[\centering Original]{%
    \label{img-orig-fullproj}%
    \includegraphics[width=0.24\textwidth,height=4.cm]{kodim04_BigC.png}}%
    \enspace
    \subfloat[\centering No Projection]{%
    \label{img-initial-fullproj}%
    \includegraphics[width=0.24\textwidth,height=4.cm]{200_initial.png}}%
    \\
    \subfloat[\centering $\ell_{1,1}$, $S=80.40$, $RM=24$]{%
    \label{img-l11-fullproj}%
    \includegraphics[width=0.24\textwidth,height=4.cm]{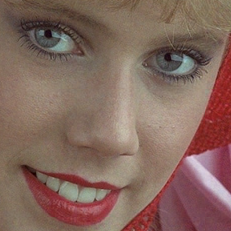}}%
    \enspace
    \subfloat[\centering $\ell_{1,\infty}$, $S=82.94$, $RM=44$]{%
    \label{img-l1inf-fullproj}%
    \includegraphics[width=0.24\textwidth,height=4.cm]{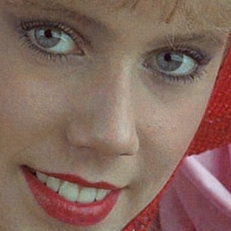}}%
    \caption{Projection of all layers : Comparison of reconstructed test images for different models. $S$ is the sparsity, $RM$ refers to the percentage of MACCS reduced by the projections. Bitrates around $2.25bpp$.}
\end{figure}

\vspace{2cm}

\section{Discussion}
The aim of this study was not to present a new compressive network with better performance than state-of-the-art networks, but rather to prove that, for any given network, the $\ell_{1,1}$ constraint and the double descent algorithm can be used as a way to efficiently and effectively reduce both storage and power consumption costs, with minimal impact on the network's performance. 

\begin{figure}[!h]
    \centering
    \includegraphics[width=0.49\textwidth,height=5.cm]{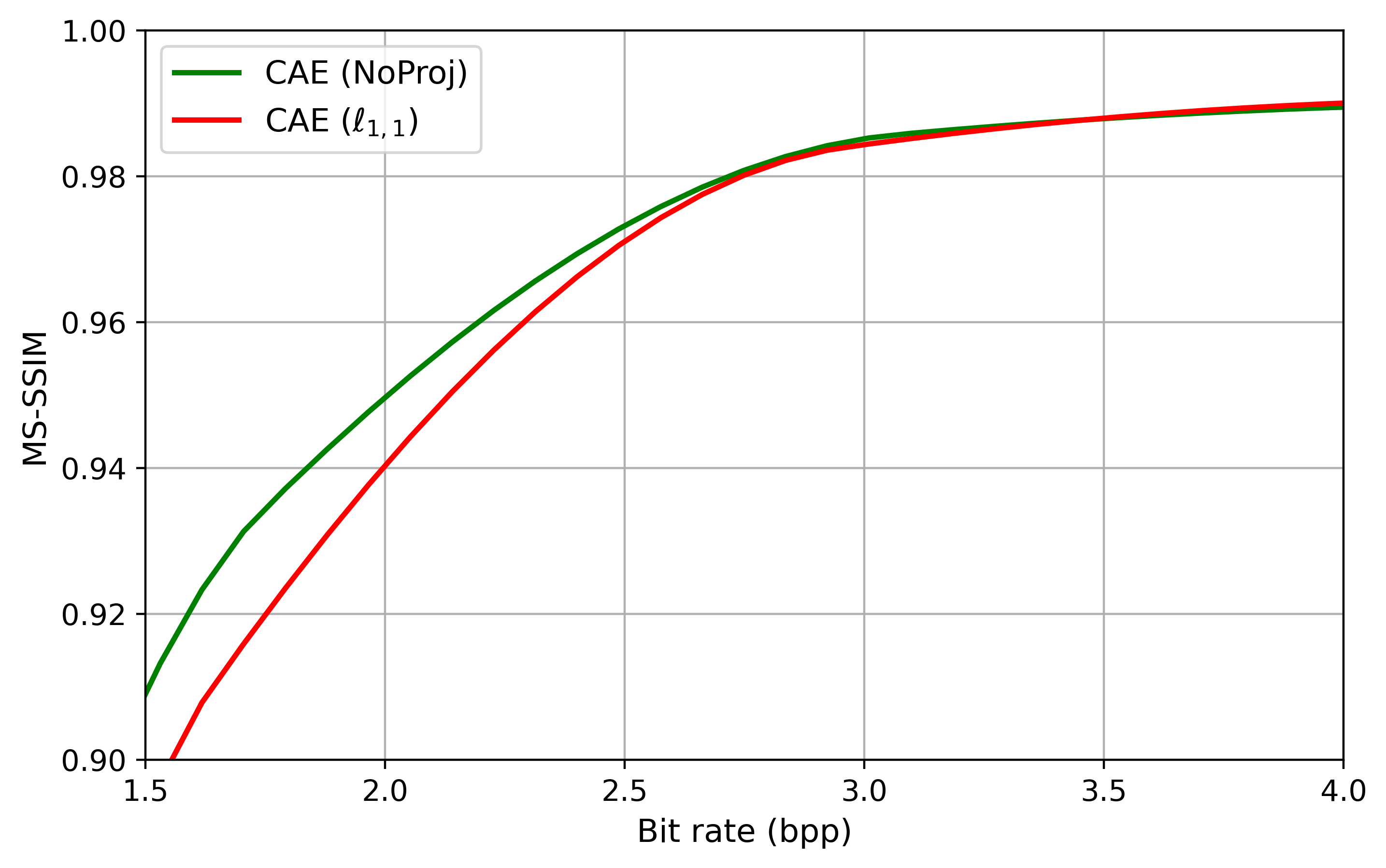}
    \caption{Mean SSIM comparison of the CAE with and without sparsification of the encoder, on Kodak.}
    \label{ssim_compar}
\end{figure}

We focus in this paper on high-quality image compression. In satellite imagery, the available energy is a crucial resource, making the image sender a prime benefactor of energy-sparing compression solutions. Meanwhile, to facilitate processing on the ground, it is also critical that the received image has the highest possible quality i.e the least possible amount of noise. The current onboard compression method on "Pleiades" CNES satellites is based on a strip wavelet transform \cite{Strip}.  Satellite imagery is a perfect application for our modified CAE method using the projection of the encoder layers, i.e. the on-board segment of the network, since it led to almost the same loss (see figure \ref{ssim_compar}) while reducing MACCS and energy consumption by 25\% and memory requirements by 82.03\% respectively.

\begin{figure}[!h]
    \centering
    \includegraphics[width=0.49\textwidth,height=5.cm]{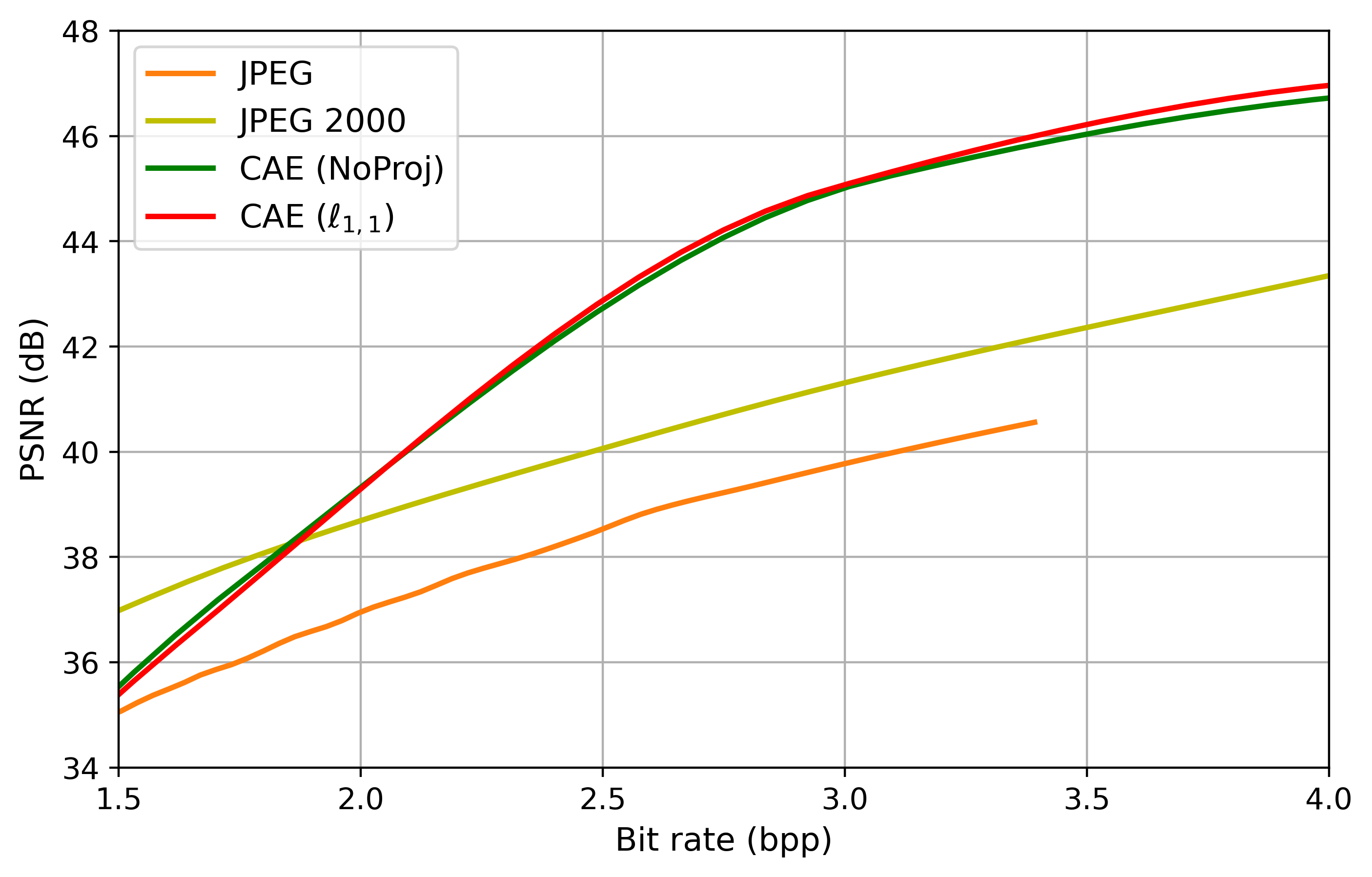}
    \caption{Comparison of the CAE with JPEG and JPEG2K at high bit-rates on Kodak. Data for JPEG and JPEG2K comes from the CompressAI benchmarks : \url{https://github.com/InterDigitalInc/CompressAI/tree/master/results/kodak}.}
    \label{high_rate}
\end{figure}

\begin{figure}[!h]
    \centering
    \includegraphics[width=0.22\textwidth]{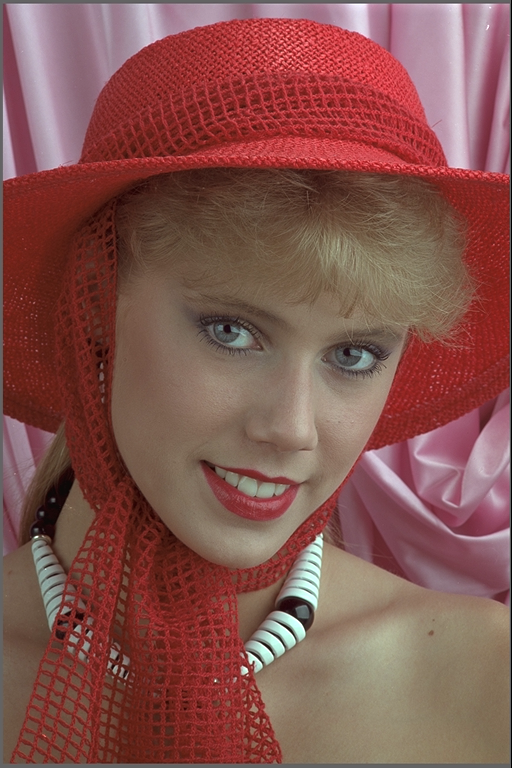}
    \includegraphics[width=0.22\textwidth]{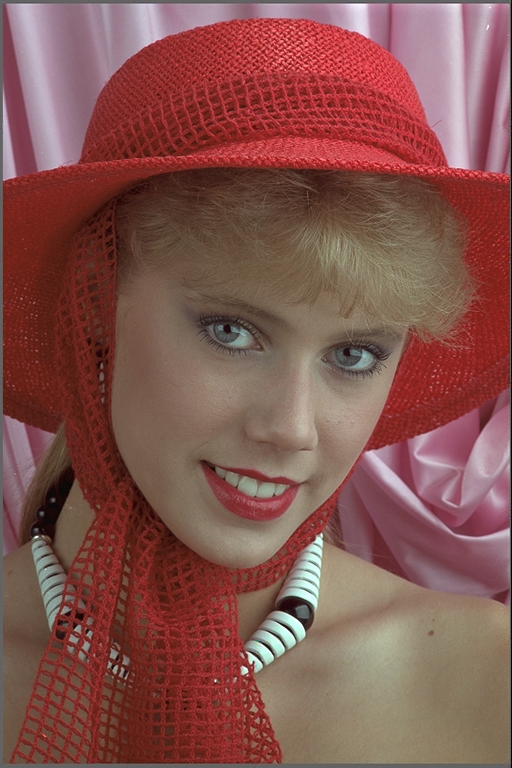}
    \caption{Kodak test images. Left: Original, Right: CAE with $\ell_{1,1}$ and $4.28$bpp.}
    \label{img-4dB-big}
\end{figure}

\begin{figure}[!h]
    \centering
    \includegraphics[width=0.22\textwidth]{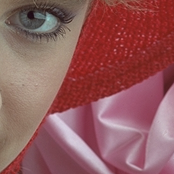}
    \includegraphics[width=0.22\textwidth]{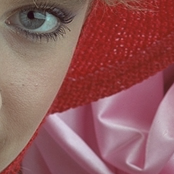}
    \caption{Kodak test images (detail). Left: Original, Right: CAE with $\ell_{1,1}$ and $4.28$bpp.}
    \label{img-4dB-zoom}
\end{figure}

Figure \ref{high_rate} shows that the SAE with low energy consumption outperforms JPEG and JPEG2K by at least $4$dB at high bit-rates (4bpp). Note that our SAE was optimized for high bit-rates: we can see the translation of the 47 dB of PSNR to unperceivable differences on test images in figures \ref{img-4dB-big} and \ref{img-4dB-zoom}. In this case (bitrate around 4bpp), we can say that we have a near lossless compression even while reducing the energy and memory costs of the network. \\

Other examples include drone imagery, media transmission from a remote or hazardous country. \\
In applications where a loss of quality is acceptable, projecting the entire network or the decoder layers can still be an efficient way to reduce a network's hardware requirements. For example, smartphones may benefit greatly from an energy-efficient compression method to save on battery life, and are not as impacted by the loss in quality that comes with the full sparsification.

This result is also very encouraging in a time where the ecological impact of neural networks is considerable \cite{AIgreen,AIred}.

\section{Conclusion and perspectives} 
\label{Conclusion}
We have proposed a framework to reduce the storage and computational cost of compressive autoencoder networks for image coding, which is crucial for several of their applications, for example mobile devices. 
Both projection $\ell_1$ and $\ell_{1,1}$ decrease the memory footprint of the network, but only the $\ell_{1,1}$ constraint decreases its computational cost without degrading the performance. \\


In this paper, we have applied the same constraint on all layers. We will adapt the constraint specifically for sparsifying either the encoder or the decoder. In future works, we will also study a layer-wise adapted constraints approach.\\ 

We have shown the interest of our method of sparsification to reduce the energy and memory costs of a CAE network. Inherently, our method is applicable to any CAE. Further works will include application of our sparsification technique to state-of-the-art compression models, such as the model from Minnen et al. \cite{toderici}, with smaller bitrates.

\appendix
\subsection*{Sparsification of the decoder layers}

In this appendix we also provide the study of the sparsification of only the decoder layers.
Table \ref{loss-decodproj} shows the relative loss of the different models for a given value of $S$ around $86\%$. The table displays a slightly greater disparity between the $\ell_{1,1}$ and $\ell_1$ constraints, with $\ell_{1,1}$ still being the only one of those two constraints to offer MACCs reduction. 
The 30\% difference of computational cost reduction between $\ell_{1, \infty}$ and $\ell_{1,1}$ remains, with the noticeable loss in performance for $\ell_{1, \infty}$.

\begin{table}[!ht]
    \centering
    \caption{Projection on decoder layers: Sparsity, MACCs and relative loss}
    \begin{tabular}{|c|c|c|c|}
        \hline
        Projection   & $\ell_{1} $  & $\ell_{1,1} $  & $\ell_{1,\infty}$ \\
        \hline
        $S$ (\%) & 86.27 & 87.40  & 87.07\\
        \hline
        MACCs Reduction $ \% $  & 0 & 33 & 61\\
        \hline
        Memory Reduction $ \% $  & 86 & 87 & 86\\
        \hline
        Relative Loss (dB)  & -4.02 & -4.32 & -8.52\\
        \hline
    \end{tabular}
    \label{loss-decodproj}
\end{table}

\begin{figure}[!ht]
    \centering
    \includegraphics[width=0.49\textwidth,height=5.cm]{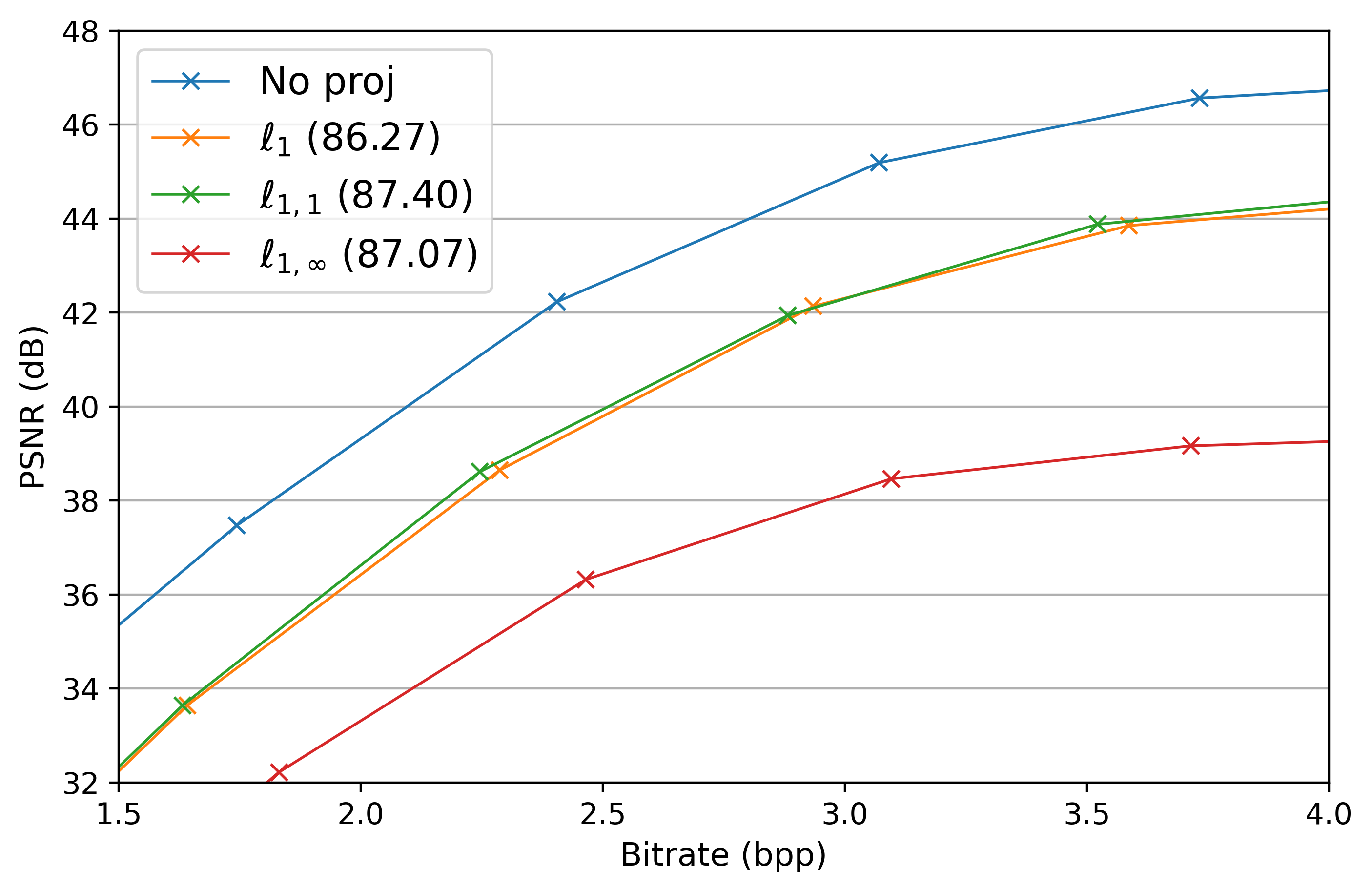}
    \caption{Projection on decoder layers: PSNR as a function of the bitrate for decoded Kodak test images}
    \label{PSNR-decodproj}
\end{figure}

\begin{figure}[!ht]
    \centering
    \includegraphics[width=0.49\textwidth,height=5.cm]{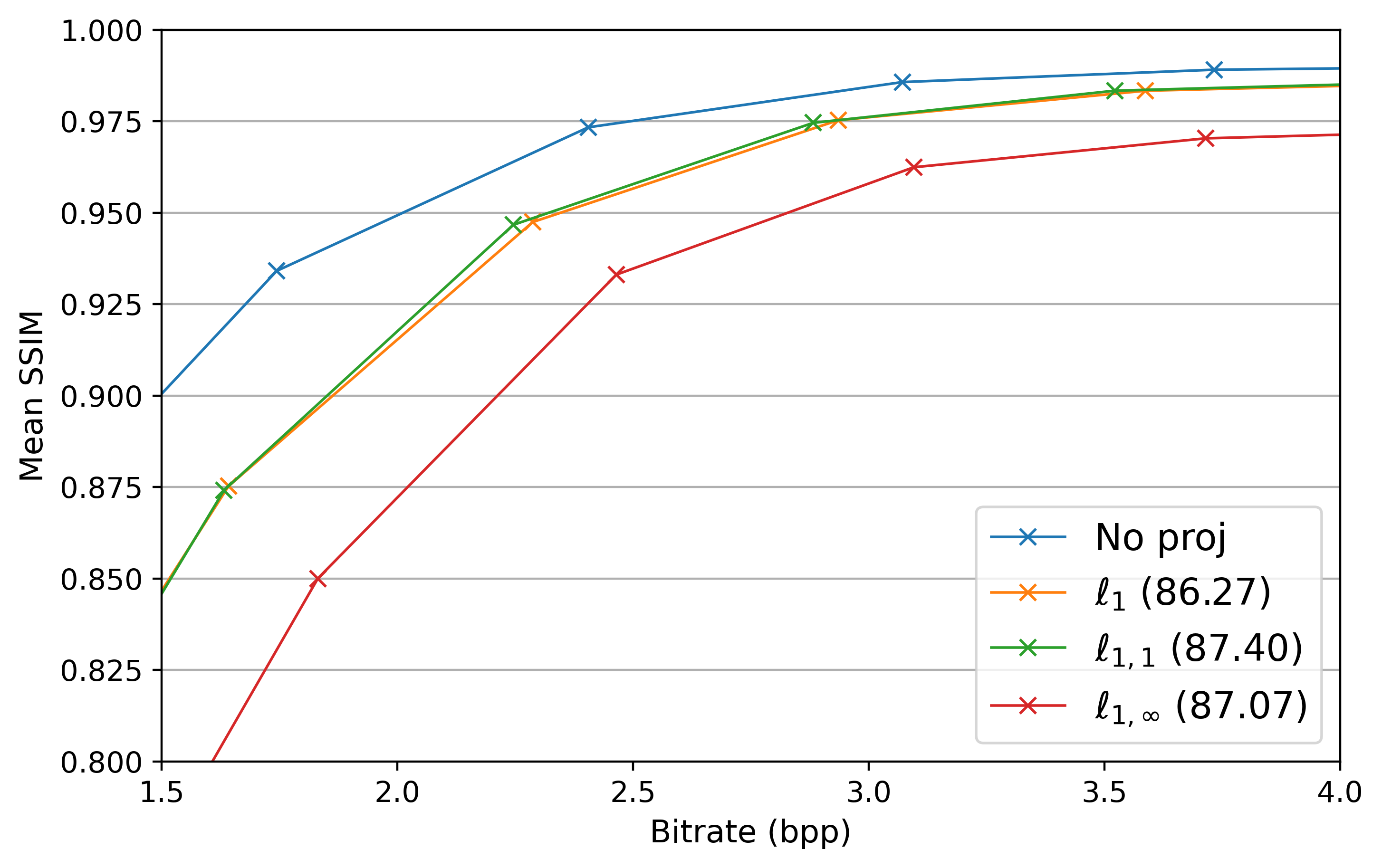}
    \caption{Projection on decoder layers : Mean SSIM as a function of the bitrate for decoded Kodak test images}
    \label{SSIM-decodproj}
\end{figure}

Figures \ref{PSNR-decodproj} and \ref{SSIM-decodproj} show that interestingly, despite the reduction in computational cost provided by the $\ell_{1,1}$ constraint, the reconstructed images are similar in quality to the ones obtained with the $\ell_1$ constraint. The $\ell_{1, \infty}$ constraint still performs worse than the other projections.

The results are very similar to the ones with sparsification of the whole network.

\section*{References}
\bibliographystyle{IEEEtran}
\bibliography{references}

\end{document}